\documentclass[11pt]{article}
\usepackage{graphicx}

\usepackage{amssymb}
\usepackage[margin=1.165in]{geometry}
\usepackage{amsfonts}
\usepackage{ragged2e}
\usepackage{multirow}
\usepackage{amsmath}
\usepackage{physics}
\usepackage{enumitem}
\usepackage{textcomp}
\usepackage{setspace}
\usepackage{tabularx}
\usepackage{datetime2}
\usepackage{authblk}
\usepackage[dvipsnames]{xcolor}
\usepackage{bm}
\usepackage{orcidlink}
\usepackage{soul}
\usepackage{tcolorbox}

\usepackage[
backend=biber,
style=numeric-comp,
sorting=none
]{biblatex}
\usepackage{chngcntr}
\usepackage{tikz}
\usepackage{comment}

\usepackage{hyperref}
\hypersetup
{
colorlinks=true,
linkcolor=blue,
urlcolor={winered},
filecolor={winered},
citecolor={winered},
allcolors={blue}
}

\definecolor{plotgreen}{RGB}{20, 164, 28}
\definecolor{plotred}{RGB}{195, 7, 24}

\definecolor{desred}{RGB}{231,101,109}
\definecolor{desblue}{RGB}{105,105,253}
\definecolor{desgreen}{RGB}{120, 213, 124}
\definecolor{despurp}{RGB}{163, 73, 164}

\title{\bf{Regular Spacetimes in the \\ Effective 
Metric Description}}

\date{}
\author[1]{Mattia Damia Paciarini\footnote{{\Large\orcidlink{0009-0000-1044-341X}}\hspace{0.1cm}\raisebox{0.9ex}{\href{mailto:damiapaciarinim@qtc.sdu.dk}{damiapaciarinim@qtc.sdu.dk}}}}

\author[1]{Mikołaj Myszkowski\footnote{{\Large\orcidlink{0000-0002-5207-4509}}\hspace{0.1cm}\raisebox{0.9ex}{\href{mailto:mikolaj@qtc.sdu.dk}{mikolaj@qtc.sdu.dk}}}}

\author[1,2,3,4]{Francesco Sannino\footnote{{\Large\orcidlink{0000-0003-2361-5326}}\hspace{0.1cm}\raisebox{0.9ex}{\href{mailto:sannino@qtc.sdu.dk}{sannino@qtc.sdu.dk}}}}

\author[1]{Vania~Vellucci\footnote{{\Large\orcidlink{0000-0003-4516-0940}}\hspace{0.1cm}\raisebox{0.9ex}{\href{mailto:vellucci@qtc.sdu.dk}{vellucci@qtc.sdu.dk}}}}

\affil[1]{\small Quantum Theory Center (${\hbar}$QTC) \& D-IAS, IMADA at Southern Denmark Univ., Campusvej 55, 5230 Odense M, Denmark}

\affil[2]{\small Scuola Superiore Meridionale, Largo S. Marcellino, 10, 80138 Napoli NA, Italy}
\affil[3]{Dept. of Physics E. Pancini, Universit`a di Napoli Federico II, via Cintia, 80126 Napoli, Italy}
\affil[4]{INFN sezione di Napoli, via Cintia, 80126 Napoli, Italy}

\begin{document}
\maketitle
\begin{abstract}
Over the past few decades, significant effort has been directed towards developing various regularized models for compact objects. Recently, a new observable-based parametrization was introduced to account for black hole deformations in a model-independent fashion, referred to as the Effective Metric Description (EMD) \cite{Binetti:2022xdi, DelPiano:2023fiw, DelPiano:2024gvw, DelPiano:2024nrl}. In this paper, we carry out a systematic investigation of the regularity of static and spherically symmetric spacetimes by extending the EMD. We derive the conditions under which curvature scalars remain finite everywhere and the associated spacetime is geodesically complete. Our results are then used to classify the cores of regular spacetimes based on their asymptotic behavior near the origin. To illustrate our findings, we apply our regularity conditions to known black hole examples. 
\end{abstract}
\newpage
\tableofcontents

\section{Introduction}\label{Sec:1}
The inception of general relativity in 1915 by A. Einstein marked a turning point in our understanding of gravity \cite{Einstein1915,Will:2014kxa,Pinochet:2020obo, Jia:2023sef,Hartle:2003yu,Misner:1973prb}. Nevertheless, despite its success in explaining various experimental phenomena, it soon became clear that the theory is plagued by singularities and is therefore incomplete. In fact, the first non-trivial solution to Einstein's field equations, discovered by Karl Schwarzschild in 1916, represents a black hole with a singular origin \cite{Schwarzschild:1916uq}. Further work by R. Penrose, S. Hawking, and others led to a classification of possible singularities and solidified their status as a robust prediction of general relativity \cite{Penrose:1964wq,Hawking:1965mf, Hawking:1970zqf,Ellis:1977pj,Wald:1997wa,Penrose:1969pc,Oppenheimer:1939ue,Lifshitz:1963ps,Geroch:1968ut,Israel:1966rt}.\par
Subsequently, much effort has been directed towards modifying classical metrics aimed at eliminating the undesired singularities \cite{Frolov:2016pav, Lan:2023cvz, Cadoni:2022chn, Dymnikova:2004qg, Hayward:2005gi, Bueno:2024dgm, Bambi:2023try}. These encompass the models of black holes that are regular everywhere, including at the origin \cite{bardeen1968, Dymnikova:1992ux, Simpson:2019mud, Estrada:2024uuu}. This top-down approach is, however, by construction, model-dependent. Therefore, it is desirable to develop a framework in which corrections to classical black holes are encoded in a model-independent fashion. Additionally, we require the framework to be frame-independent, ensuring that the results are not tied to a given coordinate system. Both requirements are satisfied by parameterizing the corrections to the metric in terms of  physical quantities. This idea was explored in \cite{Binetti:2022xdi, DelPiano:2023fiw, DelPiano:2024gvw, DelPiano:2024nrl, DelPiano:2024nrl}, where such a model-independent formalism was constructed by taking advantage of the rotational symmetry and the asymptotic behavior of the system. Within this Effective Metric Description (EMD), it is possible to derive regularity conditions for Ricci and Kretschmann scalars at the horizon and express the temperature of the modified black hole in a model-independent way.\par
One of the goals of this work is a systematic investigation of the regularity of modified static and spherically symmetric metrics within the EMD framework. We begin with a brief introduction to the EMD formalism and a summary of previous results. Subsequently, we derive the necessary and sufficient EMD conditions for the metrics to be regular everywhere, including at the origin. This, in turn, allows us to classify the types of black holes based on their asymptotic behavior near the core. Finally, we analyze the geodesic completeness of the modified spacetime \cite{Hawking:1973uf}, which leads to further constraints on the modified metrics within the EMD framework. The versatility of our approach is demonstrated by applying the regularity conditions to known black hole models. As expected, we recover the previously derived results.\par
The paper is organized as follows. In Sec.~\ref{sec:2}, we summarize the EMD framework. Sec.~\ref{sec:3} is dedicated to the local expansion and derivation of the proper distance (the chosen physical observable) at a generic point in spacetime. Regularity and geodesic completeness are investigated in Sec.~\ref{sec:4}, while the results are tested against various regular and singular black hole metrics in Sec.~\ref{sec:5}. We offer our conclusions and ideas for future investigations in Sec.~\ref{Conclusion}. Further technical details can be found in the appendices.

\section{The EMD in a Nutshell}\label{sec:2}
The aim of the EMD is to parameterize corrections to the classical metric in a model-independent way and from an observer's point of view. This implies writing corrections to the metric in terms of physical observables. The approach encompasses modifications to black hole physics that respect classical symmetries. Furthermore, we assume the existence of a metric tensor that preserves the classical asymptotic behavior. The EMD framework for static and spherically symmetric objects was developed in  \cite{Binetti:2022xdi,DelPiano:2023fiw}, and can be parameterized in the following way:
\begin{equation}\label{metric}
    ds^2=-h(r) dt^2+\frac{dr^2}{f(r)}+r^2 d\Omega_2^2, \ \ \ \ \ \ d\Omega_2^2=d\theta^2+\sin^2(\theta)d\phi^2,
\end{equation}
where ${h(r)}$ and ${f(r)}$ are the metric functions encoding the information about the deformations.\par 
At the classical level, these functions are constrained by the vacuum Einstein field equations, and the solution reduces to
\begin{equation}\label{fhclass}
    h(r)=f(r)=1-\frac{2M}{r}.
\end{equation}
For modified black holes, however, the Einstein field equations are no longer assumed to hold. Therefore, in order to account for the possible deformations of the classical geometry, we introduce the modified metric functions
\begin{equation}\label{fh}
    h(r)=1-\frac{2M}{r}\Psi(\mathcal{O}_{bs}),  \ \ \ f(r)=1-\frac{2M}{r}\Phi(\mathcal{O}_{bs}),
\end{equation}
where $\Phi(\mathcal{O}_{bs})$ and $\Psi(\mathcal{O}_{bs})$ are not generic parts of the metric functions, but rather physical quantities, since they can be defined in a coordinate-invariant way (see Appx.~\ref{Appx:C}).\par 
In the EMD, $\mathcal{O}_{bs}$ is chosen to be a physical observable \cite{DelPiano:2024gvw}. A well-suited choice for spherically symmetric spacetimes is the radial proper distance from the origin
\begin{equation}\label{propdist}
    \mathcal{O}_{bs}\rightarrow d(r)=\int_{0}^{r}\frac{dz}{\sqrt{|f(z)|}},
\end{equation}
which ensures that ${\mathcal{O}_{bs}=d}$ is a monotonically increasing function of $r$. This, in turn, guarantees that the deformation functions ${\Phi(d),\Psi(d)}$ can be chosen to represent the desired metric. We impose that, asymptotically, the metric reduces to that of Schwarzschild:
\begin{equation}
    \lim_{d\rightarrow\infty} \Phi(d)= \lim_{d\rightarrow\infty}\Psi(d)=1.
\end{equation}
When the metric function $f(r)$ has a double zero, the proper distance \eqref{propdist} diverges and a more suitable physical quantity is needed\footnote{This situation might occur in regular black hole models for particular values of the metric parameters, for which the spacetime presents an extremal horizon. In that case, the proper distance can be replaced by an alternative well-behaved physical quantity \cite{DelPiano:2024gvw,DamiaPaciarini:2025xjc}.}.\par
The metric \eqref{metric} with \eqref{fh} can subsequently be used to examine the deformed black hole in the vicinity of the outer horizon \cite{Binetti:2022xdi, DelPiano:2023fiw}, located at $d_H=d(r_H)$. Assuming that ${\Phi(d),\Psi(d)}$ admit a Taylor expansion near the horizon
\begin{equation}\label{correxp}
\begin{split}
&\Phi(d)=\Phi_H^{(0)}+\Phi^{(1)}_H(d-d_H)+\frac{1}{2}\Phi^{(2)}_H(d-d_H)^2+\mathcal{O}((d-d_H)^3),\\
&\Psi(d)=\Psi_H^{(0)}+\Psi^{(1)}_H(d-d_H)+\frac{1}{2}\Psi^{(2)}_H(d-d_H)^2+\mathcal{O}((d-d_H)^3),
\end{split}
\end{equation}
the authors were able to derive coordinate-invariant conditions for the Ricci and Kretschmann scalars to be finite there
\begin{equation}
    \Phi_H^{(1)}=\Phi_H^{(3)}=\Psi_H^{(1)}=\Psi_H^{(3)}=0.
\end{equation}
The expansion coefficients in \eqref{correxp} can be used to parameterize various properties of a modified black hole \cite{DelPiano:2024nrl, DelPiano:2023fiw}. For example, the Hawking temperature
\begin{equation}
    T=\frac{1}{4 \pi} \sqrt{\frac{1+\varpi}{2 r_H^2}-2\frac{ \Psi_H^{(2)}}{\Psi_H^{(0)}}}, \ \ \text{where} \ \ \ \varpi=\sqrt{1-\frac{8 r_H^2 \Phi_H^{(2)}}{\Phi_H^{(0)}}},
\end{equation}
can be expressed solely in terms of the second-order derivatives of ${\Phi(d),\Psi(d)}$ and the value of the radial coordinate $r_H$ at the horizon\footnote{In the original papers, the authors used a slightly different parametrization of the deformation functions involving an exponential. The formulas given here have been adjusted to conform to the parametrization \eqref{fh}.}.  

\section{Local Expansion of the Proper Distance}\label{sec:3}
In order to express the scalar invariants and other quantities of interest in terms of the deformation functions ${\Phi(d),\Psi(d)}$, we must first solve for the proper distance in equation \eqref{propdist}. This equation can be expressed in differential form as
\begin{equation}\label{propdistdiff}
\frac{\partial d(r)}{\partial r}=\frac{1}{\sqrt{|1-\frac{2M}{r}\Phi(d(r))|}}.
\end{equation}
By integrating, it is possible to obtain the local expression for $d(r)$ as a power series expansion in $r$. As we shall see, the exact form of ${d(r)}$ depends on the behavior of the deformation function ${\Phi(d)}$ near the expansion point ${r_P}$. There are three different possibilities, which must be treated separately.

\subsection{Points ${\Phi_P^{(0)}\neq \frac{r_P}{2M}, d\neq 0}$}\label{sec:3.1}
Suppose we are solving the equation \eqref{propdistdiff} in the vicinity of a generic point ${P}$ at which the first coefficient of the expansion \eqref{correxp} satisfies $\Phi_P^{(0)}\neq \frac{r_P}{2M}$, i.e., ${f(r_P)\neq 0}$. The equation for the proper distance can then be expanded in terms of the deviation of the radial coordinate ${\delta r=r-r_P}$ from the point $P$. To the leading order, the equation reads
\begin{equation}
    \frac{\partial d(r)}{\partial r}=\frac{1}{\sqrt{|1-\frac{2M}{r_P}\Phi_P^{(0)}|}}\left(1+\mathcal{O}(\delta r)\right)=\frac{1}{\sqrt{|f(r_P)|}}+\mathcal{O}(\delta r),
\end{equation}
and can be integrated to give
\begin{equation}
    d(r)=d(r_P)+\frac{1}{\sqrt{|1-\frac{2M}{r_P}\Phi_P^{(0)}|}}\delta r+\mathcal{O}(\delta r^2).
\end{equation}
Repeating this procedure order by order, we arrive at the local expression for the proper distance around $r_P$ in the form
\begin{equation}
   d(r)=d(r_P)+d_1\delta r+d_2\delta r^2+\mathcal{O}(\delta r^3),
\end{equation}
where
\begin{equation}
    d_1=\frac{1}{\sqrt{|1-\frac{2M}{r_P}\Phi_P^{(0)}|}}, \ \ \ d_2=\frac{M\left(r_P\Phi_P^{(1)}-\Phi_P^{(0)}\sqrt{|f(r_P)|}\right)}{2r_P^2|f(r_P)|f(r_P)}.
\end{equation}
These expressions can subsequently be substituted into \eqref{fh} to evaluate the expansions of the metric functions ${h(r),f(r)}$ in the neighborhood of $P$:
\begin{equation}
\begin{split}
    h(r)&=1-\frac{2M}{r_P}\Psi_P^{(0)}+\frac{2M}{r_P^2}\left(\Psi_P^{(0)}-d_1 r_P \Psi_P^{(1)}\right)\delta r\\
    &-\frac{M}{r_P^3}\left(2\Psi_P^{(0)}-2d_1 r_P \Psi_P^{(1)}+d_2 r_P^2\Psi_P^{(1)}+d_1^2 r_P^2 \Psi_P^{(2)}\right)\delta r^2+\mathcal{O}(\delta r^3),\\
    f(r)&=1-\frac{2M}{r_P}\Phi_P^{(0)}+\frac{2M}{r_P^2}\left(\Phi_P^{(0)}-d_1 r_P \Phi_P^{(1)}\right)\delta r\\
    &-\frac{M}{r_P^3}\left(2\Phi_P^{(0)}-2d_1 r_P \Phi_P^{(1)}+d_2 r_P^2\Phi_P^{(1)}+d_1^2 r_P^2 \Phi_P^{(2)}\right)\delta r^2+\mathcal{O}(\delta r^3).
\end{split}
\end{equation}
These expansions of the metric functions at different points prove particularly useful when evaluating local quantities, such as curvature scalars. We will use them in Sec.~\ref{sec:4} to derive regularity conditions for a modified black hole.

\subsection{The origin ${d=0}$}\label{sec:3.2}
The situation is different at the origin, where the metric functions potentially diverge due to the ${1/r}$ factor in \eqref{fh}. Again, we begin by expanding the differential equation for the proper distance in powers of $\delta r$. Suppose that, to leading order, ${d=d_1 \delta r^{k}+...,k>0}$. After expanding, we obtain
\begin{equation}\label{expeq0}
    d_1 k\delta r^{k-1}+...=\frac{1}{\sqrt{|1-2M\Phi_0^{(0)}\delta r^{-1}-2M\Phi_0^{(1)}d_1\delta r^{k-1}+...|}}.
\end{equation}
Depending on the value of ${\Phi_0^{(0)}}$, either the $\mathcal{O}(1)$ term or the $\mathcal{O}(\delta r^{-1})$ term on the right-hand side of \eqref{expeq0} dominates close to the origin.\par
\subsubsection* {\bf Case 1: $\Phi_0^{(0)}=0$} \par
If $\Phi_0^{(0)}=0$, then it is the $\mathcal{O}(1)$ term that dominates the denominator of \eqref{expeq0} close to the origin, and ${k=1}$. Solving for higher-order terms leads to
\begin{equation}
    d(r)=d_1 \delta r+d_2 \delta r^2+d_3 \delta r^3+\mathcal{O}(\delta r^4).
\end{equation}
which, after substituting into \eqref{fh}, results in the following expansions of the metric functions:
\begin{equation}\label{metrexp0}
\begin{split}
h(r)=&1-2Md_1\Psi_0^{(1)}-M\left(2d_2 \Psi_0^{(1)}+d_1^2\Psi_0^{(2)}\right)\delta r\\
&-2M\left(d_3\Psi_0^{(1)}+d_1d_2\Psi_0^{(2)}+\frac{1}{6}d_1^3\Psi_0^{(3)}\right)\delta r^2+\mathcal{O}(\delta r^3),\\
f(r)=&1-2Md_1\Phi_0^{(1)}-M\left(2d_2 \Phi_0^{(1)}+d_1^2\Phi_0^{(2)}\right)\delta r\\
&-2M\left(d_3\Phi_0^{(1)}+d_1d_2\Phi_0^{(2)}+\frac{1}{6}d_1^3\Phi_0^{(3)}\right)\delta r^2+\mathcal{O}(\delta r^3).
\end{split}
\end{equation}
The expressions for the first two coefficients ${d_1,d_2}$, relevant to our calculations, are given in Appx.~\ref{Appx:A}.

\subsubsection*  {\bf Case 2: $\Phi_0^{(0)}\neq 0$}\par
If $\Phi_0^{(0)}\neq 0$, then the ${-2M\Phi_0^{(0)}\delta r^{-1}}$ term dominates on the right-hand side of equation \eqref{expeq0}, which implies $k=3/2$. Integrating order by order, the proper distance takes the form
\begin{equation}\label{propdistd0case2}
    d(r)=d_1 \delta r^{3/2}+d_2 \delta r^{5/2}+d_3 \delta r^{3}+\mathcal{O}(\delta r^{7/2}),
\end{equation}
where, unlike previously, extra terms with non-integer powers of $\delta r$ appear in the expansion. The origin of these non-integer powers can be traced back to the divergence of ${f(r)}$ at ${r=0}$. The expressions for the coefficients ${d_1,d_2,d_3}$ are given in Appx.~\ref{Appx:A}.\par
The proper distance \eqref{propdistd0case2} can then be substituted into \eqref{fh}, which gives the expansions of ${h(r),f(r)}$ in terms of $\delta r$:
\begin{equation}\label{metrexp00}
\begin{split}
h(r)=&-\frac{2M\Psi_0^{(0)}}{\delta r}+1-2Md_1\Psi_0^{(1)}\delta r^{1/2}\\
&-2M d_2\Psi_0^{(1)}\delta r^{3/2}-M\left(2d_3\Psi_0^{(1)}+d_1^2\Psi_0^{(2)}\right)\delta r^2+\mathcal{O}(\delta r^{5/2}),\\
f(r)=&-\frac{2M\Phi_0^{(0)}}{\delta r}+1-2Md_1\Phi_0^{(1)}\delta r^{1/2}\\
&-2M d_2\Phi_0^{(1)}\delta r^{3/2}-M\left(2d_3\Phi_0^{(1)}+d_1^2\Phi_0^{(2)}\right)\delta r^2+\mathcal{O}(\delta r^{5/2}).
\end{split}
\end{equation}

\subsection{Points $\Phi_P^{(0)}=\frac{r_P}{2M}, d\neq 0$}\label{sec:3.3}
The third special case occurs at points $\Phi_P^{(0)}=\frac{r_P}{2M},d\neq 0$, where ${f(r_P)=0}$. We again start by assuming that, to leading order, ${d=d(r_P)+d_1\delta r^{k}+...,k>0}$. Expanding the denominator of \eqref{propdistdiff} in powers of ${\delta r}$, the first two terms cancel out, allowing us to write
\begin{equation}\label{diffeq3}
 d_1 k\delta r^{k-1}+...=\frac{1}{\sqrt{|\frac{\delta r}{r_P}-\frac{2M}{r_P}\Phi_P^{(1)}d_1\delta r^k-\frac{M}{r_P}\Phi_P^{(2)}d_1^2\delta r^{2k}+...|}}.
\end{equation}
Therefore, the leading power $k$ depends on whether ${\Phi_P^{(1)}}$ vanishes.\par
\subsubsection*  {\bf Case 1: $\Phi_P^{(1)}=0$}\par
If $\Phi_P^{(1)}=0$, the term $\delta r/r_P$ dominates the denominator of \eqref{diffeq3}, resulting in $k=1/2$. Since ${f(r)}$ is zero at $r_P$, the absolute value in \eqref{diffeq3} necessitates separate expansions for $\delta r\geq0$ and $\delta r\leq0$. The two expressions can then be sewn together at $r=r_P$, resulting in
\begin{equation}\label{propdist4}
d(r)= 
    \begin{cases}
      d(r_P)+d_1^{+}|\delta r|^{1/2}+d_2^{+}|\delta r|+\mathcal{O}(|\delta r|^{3/2}) & \text{for} \ \  \delta r\geq0\\
      d(r_P)+d_1^{-} |\delta r|^{1/2}+d_2^{-} |\delta r|+\mathcal{O}(|\delta r|^{3/2}) & \text{for} \ \  \delta r\leq0
    \end{cases}
\end{equation}
where the constants ${d_1^{\pm},d_2^{\pm}}$, expressed in terms of the deformation functions $\Phi(d),\Psi(d)$, are given in Appx.~\ref{Appx:A}. Using \eqref{fh}, we obtain the expansion of the metric functions $h(r),f(r)$ in the vicinity of $r_P$:
\begin{equation}\label{exphor1}
\begin{split}
h(r)=&1-\frac{2M}{r_P}\Psi_P^{(0)}-\frac{2M d_1^{\pm}\Psi_P^{(1)}}{r_P}|\delta r|^{1/2}\\
&+\frac{M}{r_P^2}\left(\pm 2\Psi_P^{(0)}-2 d_2^{\pm} r_P\Psi_P^{(1)}-(d_1^{\pm})^2 r_P \Psi_P^{(2)}\right)|\delta r|+\mathcal{O}(|\delta r|^{3/2}),\\
f(r)=&\frac{1\mp M (d_1^{\pm})^2\Phi_P^{(2)}}{r_P}|\delta r|-\frac{M}{3r_P}\left(6d_1^{\pm} d_2^{\pm} \Phi_P^{(2)}+(d_1^{\pm})^3\Phi_P^{(3)}\right)|\delta r|^{3/2}+\mathcal{O}(|\delta r|^2),
\end{split}
\end{equation}
where ${+}$ and ${-}$ correspond to the regions $\delta r\geq 0$ and $\delta r \leq0$, respectively.\par

\subsubsection*  {\bf Case 2: $\Phi_P^{(1)}\neq 0$}\par

If $\Phi_P^{(1)}\neq 0$, it can be shown that $-\frac{2M}{r_P}\Phi_P^{(1)}d_1\delta r^k$ is the leading term in the denominator of \eqref{diffeq3} when $\delta r$ is small. Indeed,
solving $\eqref{diffeq3}$ to the leading order in $\delta r$ gives $k=2/3$ and the term $-\frac{2M}{r_P}\Phi_P^{(1)}d_1\delta r^{2/3}$ dominates over $\delta r/r_P$. As in the case $\Phi_P^{(1)}=0$, we consider the regions $\delta r\geq 0$ and $\delta r\leq 0$ separately. Repeating this process at higher orders and matching the coefficients, we arrive at the following expansions of the proper distance:
\begin{equation}\label{propdist5}
d(r)= 
    \begin{cases}
      d(r_P)+d_1^{+}|\delta r|^{2/3}+d_2^{+}|\delta r|+\mathcal{O}(|\delta r|^{4/3}) & \text{for} \ \  \delta r\geq 0\\
      d(r_P)+d_1^{-} |\delta r|^{2/3}+d_2^{-} |\delta r|+\mathcal{O}(|\delta r|^{4/3}) & \text{for} \ \  \delta r\leq0
    \end{cases}    
\end{equation}
Using \eqref{propdist5} and \eqref{fh}, the expansions of the metric functions in the neighborhood of $r_P$ can be written as
\begin{equation}
\begin{split}
&h(r)=1-\frac{2M}{r_P}\Psi_P^{(0)}-\frac{2Md_1^{\pm}\Psi_P^{(1)}}{r_P}|\delta r|^{2/3}\pm\frac{2M}{r_P^2}\left(\Psi_P^{(0)}-d_2^{\pm}r_P\Psi_P^{(1)}\right)|\delta r|+\mathcal{O}(|\delta r|^{4/3}),\\
&f(r)=-\frac{2Md_1^{\pm}\Phi_P^{(1)}}{r_P}|\delta r|^{2/3}\pm\frac{1-2Md_2^{\pm}\Phi_P^{(1)}}{r_P}|\delta r|+\mathcal{O}(|\delta r|^{4/3}).
\end{split}
\end{equation}
The expressions for the coefficients ${d_1^{\pm},d_2^{\pm}}$ are once again given in Appx.~\ref{Appx:A}.\par
Although the intuition behind these calculations is rather elementary, there are important technical details that we have chosen to omit here. A more complete discussion of these technicalities can be found in Appx.~\ref{Appx:B}.

\section{Regularity and Properties of the Deformed Spacetime}\label{sec:4}
Following the derivation of the proper distance and metric expansions in the vicinity of an arbitrary point, we are now ready to examine the regularity of the spacetime described by generic deformation functions ${\Phi(d),\Psi(d)}$. Since there is no universal definition of regularity for generic spacetimes, we focus on the two aspects most commonly discussed in the literature: curvature scalars and geodesic completeness.
\subsection{Finiteness of curvature scalars}\label{sec:4.1}
We begin by considering the finiteness of the Kretschmann curvature scalar $K$. In spacetimes that are both static and spherically symmetric, the Kretschmann scalar is a sum of squares of all nonzero components of the Riemann tensor. For this reason, it can be demonstrated that in such spacetimes, the Kretschmann scalar is finite if and only if all components of the Riemann tensor, evaluated in an orthonormal frame, are themselves finite. Therefore, in a static and spherically symmetric spacetime, the finiteness of the Kretschmann scalar is sufficient to guarantee that all scalar polynomials constructed from the Riemann tensor (together with the metric and the fully antisymmetric Levi-Civita tensor) are also finite \cite{Bronnikov:2012wsj}.\par
The metric expansions developed in Sec.~\ref{sec:3} can now be substituted one by one into the expression for $K$. Requiring the finiteness of $K$ (and, by extension, all scalar polynomials constructed from the Riemann tensor) at a generic point in spacetime then leads to non-trivial regularity conditions, which are summarized below.
\begin{tcolorbox}
\begin{center}
{\bf Finiteness of $K$ at points ${\Phi_P^{(0)}\neq 
\frac{r_P}{2M}, d\neq 0}$:} \par
\vspace{0.05cm}
$\Psi_P^{(0)}\neq \frac{r_P}{2M}$\par
\end{center}
\end{tcolorbox}

\begin{tcolorbox}
\begin{center}
{\bf Finiteness of $K$ at the origin $d=0$:} \par
\vspace{0.1cm}
$\Phi_0^{(0)}=\Phi_0^{(1)}=\Phi_0^{(2)}=0$, \ \ \ 
$\Psi_0^{(0)}=\Psi_0^{(2)}=0, \ \Psi_0^{(1)}\neq \frac{1}{2M}$
\end{center}
\end{tcolorbox}

\begin{tcolorbox}
\begin{center}
{\bf Finiteness of $K$ at points ${\Phi_P^{(0)}=\frac{r_P}{2M}, d\neq 0}$:} \par
\vspace{0.1cm}
$\Phi_P^{(1)}=0,\Psi_P^{(0)}\neq \frac{r_P}{2M}$, or\par 
\vspace{0.1cm}
If $\frac{d}{dr}f(r)>0$:\par
\vspace{0.1cm}
$\Psi_P^{(0)}=\frac{r_P}{2M}, \ \Phi_P^{(1)}=\Psi_P^{(1)}=0, \ \Psi_P^{(2)}\neq \pm\frac{1}{M(d_1^{\pm})^2}, \  \Phi_P^{(2)}\Psi_P^{(3)}=0 , \ \Phi_P^{(3)}+2\Psi_P^{(3)}=0$\par
\vspace{0.1cm}
If $\frac{d}{dr}f(r)<0$:\par 
\vspace{0.15cm}
$\Psi_P^{(0)}=\frac{r_P}{2M}, \ \Phi_P^{(1)}=\Psi_P^{(1)}=0, \ \Psi_P^{(2)}\neq \pm\frac{1}{M(d_1^{\pm})^2}, \  \Phi_P^{(2)}\Psi_P^{(3)}=0, \ \Phi_P^{(3)}=\Psi_P^{(3)}$\par
\end{center}
\end{tcolorbox}
The conditions on $\Phi(d), \Psi(d)$ at the points where ${h(r_P)=f(r_P)=0}$ depend on whether $f(r)>0$ or $f(r)<0$ when $\delta r>0$, i.e., on the sign of the first derivative of $f(r)$. The above conditions are necessary and sufficient to ensure that the Kretschmann scalar and all curvature scalars constructed from the Riemann tensor remain finite everywhere. 

\subsection{Geodesic completeness}\label{sec:4.2}
An alternative criterion for the regularity of spacetimes can be formulated in terms of geodesics \cite{Hawking:1973uf}. A spacetime is said to be geodesically complete if all geodesics can be extended to infinite values of their affine parameters. This ensures that the paths of test particles do not end abruptly and their trajectories are well-defined at all times. Conversely, a spacetime is singular at a point $P$ if there exists a geodesic that reaches $P$ in a finite affine parameter and cannot be extended.\par
It should be noted that geodesic completeness and the finiteness of curvature polynomials are not equivalent. Examples exist of spacetimes with finite curvature scalars that are not geodesically complete \cite{Taub:1950ez, Newman:1963yy}. Conversely, there are also examples of geodesically complete spacetimes with diverging curvature scalars \cite{Podolsky:2004qu,Kundt1961}. In what follows, we assume that the regularity conditions imposed by the curvature scalars are already satisfied.\par
The behavior of geodesic lines is governed by a set of geodesic equations, which, for a static spherically symmetric spacetime, can be reduced to the form \cite{Hartle:2021pel}
\begin{equation}\label{geodeq}
    \left(\frac{dr}{d\lambda}\right)^2+V(r)=0, \ \ \ V(r)=\left(1-\frac{2M}{r}\Phi(d(r))\right)\left(\frac{L^2}{r^2}-\frac{E^2}{1-\frac{2M}{r}\Psi\left(d(r)\right)}-\kappa\right),
\end{equation}
where 
\begin{equation}
     E=\left(1-\frac{2M}{r}\Psi\left(d(r)\right)\right)\frac{dt}{d\lambda}, \ \ \ L=r^2\frac{d\phi}{d\lambda},
\end{equation}
are the two constants of motion corresponding to time translations and rotational symmetry, ${\lambda}$ denotes the affine parameter, and $\kappa=-1$ (or $\kappa=0$) for timelike (or null) geodesics. Using the expansions of the metric functions ${h(r),f(r)}$ developed in Sec.~\ref{sec:3}, we can Taylor approximate \eqref{geodeq} and examine the regularity of geodesics in the neighborhood of each point. If there exists at least one timelike or null geodesic that terminates at any point with a finite value of affine parameter and cannot be extended beyond that point, the spacetime is geodesically incomplete.\par
In order to examine the behavior of geodesics near a given point ${r_P}$, we can use the expressions for metric functions developed in Sec.~\ref{sec:3} to expand the effective potential $V(r)$ in powers of $r-r_P$. The leading-order term suffices to determine whether the geodesic terminates, or can be continued beyond ${r_P}$. Similarly to Sec.~\ref{sec:3}, the situation depends on the behavior of $h(r),f(r)$ near the expansion point.\par

\subsubsection* {\bf Geodesics completeness at ${\Phi_P^{(0)}\neq \frac{r_P}{2M}, d\neq 0}$} \par
If ${\Phi_P^{(0)}\neq \frac{r_P}{2M}, r_P\neq 0}$, the metric functions admit a local expansion in $\delta r$ with non-zero constant terms, i.e., ${h(r_P),f(r_P)\neq 0}$. Therefore, we can always transform the metric to locally flat coordinates, where the geodesics are (locally) straight lines. As a result, the geodesics are well-behaved in the vicinity of $r_P$, and no additional conditions are imposed on the deformation functions.\par

\subsubsection* {\bf Geodesics completeness at the origin $d=0$  } \par
 \par
First, consider the purely radial null geodesics. In the neighborhood of the origin, the effective potential ${V(r)}$ given in \eqref{geodeq} can be expanded in the form
\begin{equation}\label{kl0}
V(r)\propto \text{const.} \ \ \text{for} \ \  L=0, \kappa=0
\end{equation}
This can be substituted into \eqref{geodeq} and integrated to obtain a local expansion of radial geodesics near the origin. In the case of \eqref{kl0}, the geodesics reach the origin with a finite value of the affine parameter. We note that for ${h(0)<0}$, geodesics are trapped near the origin and focus to a point. Indeed, in this case, both families of radial null geodesics (here labeled with + for outgoing and - for ingoing) acquire a negative expansion parameter \cite{Poisson:2009pwt,Wang:2022ews}: 
\begin{equation}
    \theta_{-}=\nabla_\mu k^{\mu}_- \propto -\frac{1}{r}\sqrt{\frac{f(r)}{h(r)}} \quad \text{and} \quad \theta_{+}=\nabla_\mu k^{\mu}_+ \propto \frac{\text{sgn}(h(r))}{r}\sqrt{\frac{f(r)}{h(r)}} \quad 
\end{equation}
where $k^{\mu}=\frac{dx^{\mu}}{d \lambda}$ is the tangent vector to the considered geodesic.\par 
The regularity of curvature invariants requires $\sqrt{\frac{f(r)}{h(r)}}$ to be finite and non-zero\footnote{Note that the existence of a solution to the geodesic equation requires $\frac{f(r)}{h(r)}$ to be real.}. Therefore, if $h(0)<0$ then $\theta_{+}\to-\infty$ at $r=0$, signaling the focusing of all geodesics at this point. This means that there is no antipodal continuation of geodesics beyond $r=0$, and thus particle trajectories terminate there at a finite value of the affine parameter. Therefore, in order to allow for the continuation of radial null geodesics, we must impose ${h(0)>0}$ or, in terms of the deformation functions:
\begin{equation}\label{cond1}
    \frac{1}{2M}>\Psi_0^{(1)}.
\end{equation}
This condition, together with the finiteness conditions from Sec.~\ref{sec:4.1}, is sufficient to prevent the focusing of all geodesics at the origin. \par

\subsubsection* {\bf Geodesics completeness at a point $P$ for which  $\Phi_P^{(0)}=\frac{r_P}{2M}, d\neq 0$} \par
When the value of the deformation function at $r_P$ is ${\Phi_P^{(0)}=\frac{r_P}{2M}}$, the metric function ${f(r_P)}$ vanishes. However, if we have ${f(r_P)=0}$ but ${h(r_P)\neq0}$, the effective potential in \eqref{geodeq} changes sign, and, on at least one side of ${r_P}$, the geodesic equations admit no solution for real energies (in fact, in this case, the metric acquires a Euclidean signature). As a result, there are geodesics which cannot be continued beyond $r_P$. We can also verify that the affine parameter required to reach the point ${r_P}$ is finite. Expanding the effective potential \eqref{geodeq} and integrating, we obtain
\begin{equation}
    V(r)\propto \delta r+...\implies \lambda+c\propto\delta r^{1/2}.
\end{equation}
Therefore, for the spacetime to be geodesically complete, we require both metric functions to vanish simultaneously, i.e., ${h(r_P)=f(r_P)=0}$. In terms of the deformation functions, this condition can be expressed as
\begin{equation}\label{cond2}
    \Phi_P^{(0)}=\frac{r_P}{2M}\Leftrightarrow\Psi_P^{(0)}=\frac{r_P}{2M}.
\end{equation}\par
In the case ${h(r_P)=f(r_P)=0}$, the curvature scalar conditions from Sec.~\ref{sec:4.1} require that, in the neighborhood of $r_P$,
\begin{equation}
    f(r)\propto \delta r, \ \  h(r)\propto \delta r \implies V(r)\propto \text{const.}
\end{equation}
Therefore, the local solution to the geodesic equation reads
\begin{equation}
    \lambda+c\propto \delta r.
\end{equation}
Depending on the signs of ${h(r), f(r)}$, there may be both ingoing and outgoing geodesics, or only ingoing ones. However, the full determination of whether a modified black hole is geodesically complete requires further considerations beyond the scope of this paper\footnote{Potential geodesic incompleteness arises at the horizons with $f(r),h(r)<0$ in the positive $\delta r$ direction. A classical example is the inner horizon of the Reissner–Nordström black hole \cite{Reissner:1916cle,Weyl:1917rtf,1918KNAB...20.1238N,1921RSPSA..99..123J}. Geodesics are usually continued beyond such a horizon using analytical extensions of the original spacetime \cite{Hawking:1973uf}. However, this is beyond the scope of this paper and therefore the geodesic condition derived in this section is necessary but may not be sufficient to guarantee full geodesic completeness.}.\par
Finally, we note that the conditions \eqref{cond1} and \eqref{cond2} are not independent. In fact, since the finiteness of the curvature scalars at the origin requires $f(0)>0$ and the horizons to be simple zeros of $h(r)$, condition \eqref{cond2} automatically implies \eqref{cond1}. Therefore, geodesic completeness requires imposing one additional condition:
\begin{tcolorbox}
\begin{center}
{\bf Geodesic completeness:} \par
$\Phi_P^{(0)}=\frac{r_P}{2M}\Leftrightarrow\Psi_P^{(0)}=\frac{r_P}{2M}$\par
\end{center}
\end{tcolorbox}
Since $h(r)$ is a continuous function of $r$ and we have assumed that there are no higher-order horizons (i.e., $\frac{d}{dr}f(r)\neq0$ at the horizons), the number of zeroes of $f(r)$ determines the sign of $f(r)$ at the origin. In particular, this implies that the finiteness of curvature scalars can be achieved only if the spacetime possesses an even number of Killing horizons. This agrees with the analysis carried out in \cite{Carballo-Rubio:2019fnb}, where the authors arrived at the same conclusion using purely geometrical arguments.
\subsection{Classification of black hole cores}\label{sec:4.3}
The results from Sec.~\ref{sec:3.2} can now be used to examine the asymptotic behavior of the black hole near its core (i.e., near ${d(r)=r=0}$). For a regular black hole, the metric expansion \eqref{metrexp0} around the origin takes the form
\begin{equation}\label{expcores}
\begin{aligned}
&f(r)=1-\frac{1}{3}M\Phi_0^{(3)}r^2+\mathcal{O}(r^3),\\
&h(r)=1-2M\Psi_0^{(1)}-\frac{1}{3}M\left(\Psi_0^{(3)}+6d_3\Psi_0^{(1)}\right)r^2+\mathcal{O}(r^3).
\end{aligned}
\end{equation}
Black hole cores can be grouped into four categories, depending on the local expansion of the metric around the origin \cite{2020PhRvDGeodesicallyCompleteBH}. The classification, formulated in terms of the deformation functions ${\Phi(d),\Psi(d)}$, is as follows:\par
\begin{tcolorbox}
\begin{center}
{\bf Minkowski Core:} $\Psi_0^{(1)}=\Psi_0^{(3)}=\Phi_0^{(3)}=0$ \par
{\bf AdS Core:} $\Psi_0^{(1)}=0, \quad \Psi_0^{(3)}=\Phi_0^{(3)}<0$\par
{\bf dS Core:} $\Psi_0^{(1)}=0, \quad \Psi_0^{(3)}=\Phi_0^{(3)}>0$\par
{\bf Mixed Core:} $\Psi_0^{(1)}\neq 0 \ , \ \text{or} \quad  \Psi_0^{(3)}\neq \Phi_0^{(3)}$\par
\end{center}
\end{tcolorbox}
\noindent If the core is of the AdS, dS, or Minkowski type, we can also compute its cosmological constant, which turns out to be simply $\Lambda=M\Phi_0^{(3)}$.
\section{Examples}\label{sec:5}
In order to illustrate the versatility of our approach, we examine three examples: the Schwarzschild, Dymnikova, and Visser black holes \cite{Schwarzschild:1916uq, Dymnikova:1992ux, Simpson:2019mud}. The first is a well-known example of a singular black hole, while the other two are fully regular, including at the origin.\par
\subsection{Schwarzschild black hole}
The classical solution to the Einstein field equations, the Schwarzschild metric, is recovered in the limit
\begin{equation}
h(r)=f(r)=1-\frac{2M}{r}\implies \Phi(d)=\Psi(d)=1,
\end{equation}
and the metric function ${f(r)}$ vanishes at ${r=2M}$. Therefore, the points where ${\Phi_P^{(0)}\neq \frac{r_P}{2M}, d\neq 0}$ and $\Phi_P^{(0)}=\frac{r_P}{2M}, d\neq 0$ are located at ${r\in\mathbb{R_+}\setminus\{0,2M\}}$ and $r=2M$, respectively. It is then easy to verify that the set of conditions from Sec.~\ref{sec:4.1}  and Sec.~\ref{sec:4.2} is satisfied at ${r\in\mathbb{R_+}\setminus\{0\}}$. Nevertheless, at the origin we have 
\begin{equation}
    \Phi_0^{(0)}\neq 0,
\end{equation}
which violates the $d=0$ regularity conditions. Therefore, as expected, we recover the essential singularity which plagues the Schwarzschild metric.
\subsection{Dymnikova black hole}
The metric of the Dymnikova black hole can be written in Schwarzschild coordinates as
\begin{equation}\label{metrfundym}
    h(r)=f(r)=1-\frac{2M}{r}\left(1-\exp\left(-\frac{r^3\Lambda}{6M}\right)\right),
\end{equation}
where ${M}$ and ${\Lambda}$ are the two free parameters of the model. The horizon(s) of the Dymnikova black hole cannot be determined analytically and are therefore defined implicitly through the equation $f(r_P)=0$. Depending on the value of the parameters ${M,\Lambda}$, the model has either two, one, or no horizons.\par
Since for the Dymnikova model ${\Phi(d)=\Psi(d)}$, the curvature polynomials constructed from the Riemann tensor are automatically finite at the points ${\Phi_P^{(0)}\neq \frac{r_P}{2M}, d\neq 0}$. At points ${\Phi_P^{(0)}=\frac{r_P}{2M}}$, $d\neq 0$ (for the Dymnikova black hole ${f(r)=h(r)}$, and hence these points coincide with the horizons), the local expansion of the metric functions reads
\begin{equation}\label{expdym}
h(r)=f(r)=\left(\frac{1}{r_P}-r_P \Lambda+\frac{r_P^2 \Lambda}{2 M}\right)\delta r-\frac{\left(8 M^2-2 M r_P^5 \Lambda^2+r_P^6 \Lambda^2\right)}{8 M^2 r_P^2}\delta r^2+\mathcal{O}(\delta r^3).
\end{equation}
Since there is no ${\sim \delta r^{2/3}}$ term in the expansion of $f(r)$, we conclude that $\Phi_P^{(1)}=0$. Comparing the higher-order terms in the expansions \eqref{expdym} and \eqref{exphor1}, we read off the first few coefficients of the deformation functions at the horizon(s)
\begin{equation}\label{dymcoeffh}
\begin{split}
&\Phi_P^{(1)}=0, \ d_2^{\pm}\Phi_P^{(2)}+\frac{1}{6}\left(d_1^{\pm}\right)^2\Phi_P^{(3)}=0, \\ 
&\Psi_P^{(1)}=0, \ d_2^{\pm}\Psi_P^{(2)}+\frac{1}{6}\left(d_1^{\pm}\right)^2\Psi_P^{(3)}=0.
\end{split}
\end{equation}
In the case of two horizons 
\begin{equation}
\frac{1}{r_P}-r_P \Lambda+\frac{r_P^2 \Lambda}{2 M}\neq 0\implies \Psi_P^{(2)}\neq \pm\frac{1}{M(d_1^{\pm})^2},
\end{equation}
and the coefficients \eqref{dymcoeffh} satisfy the regularity conditions at the points ${\Phi_P^{(0)}=\frac{r_P}{2M}}$, $d\neq 0$.\par
Finally, we investigate the regularity of the Dymnikova model at the origin. We again start by expanding the metric functions \eqref{metrfundym} around $r=0$, and comparing with \eqref{metrexp0} and \eqref{metrexp00}. This allows us to retrieve the leading coefficients ${\Phi_0^{(n)},\Psi_0^{(n)}}$ of the deformation function expansion
\begin{equation}\label{dymcoeff0}
\begin{split}
&\Phi_0^{(0)}=\Phi_0^{(1)}=\Phi_0^{(2)}=0, \\ 
&\Psi_0^{(0)}=\Psi_0^{(1)}=\Psi_0^{(2)}=0.
\end{split}
\end{equation}
It is straightforward to verify that the above coefficients indeed satisfy the corresponding finiteness conditions at the origin. Since  $\Psi_0^{(1)}=0$ and $\Psi_0^{(3)}=\Phi_0^{(3)}=\frac{\Lambda}{M}>0$, we find that, in the case of the Dymnikova model, the classical singularity is replaced by a dS core with cosmological constant ${M\Phi_0^{(3)}=\Lambda}$. Since ${\Phi(d)=\Psi(d)}$ and the conditions for the regularity of curvature scalars are satisfied, the necessary conditions for geodesic completeness are also met.
\subsection{Visser black hole}
The Visser model \cite{Simpson:2019mud} is yet another example of a regular black hole. The major difference, however, is that the metric of this model is not analytic around the origin. We will see that even then, the deformation functions ${\Phi(d),\Psi(d)}$ admit a Taylor approximation around $r=0$, and the model can be treated using the Effective Metric Description.\par
The calculations for the points ${\Phi_P^{(0)}\neq \frac{r_P}{2M}, d\neq 0}$ and $\Phi_P^{(0)}=\frac{r_P}{2M}, d\neq 0$ largely follow the same route as for the Dymnikova black hole. The metric functions can be written in terms of the radial coordinate as
\begin{equation}\label{metrviss}
    h(r)=f(r)=1-\frac{2M}{r}\exp\left(-\frac{a}{r}\right),
\end{equation}
where ${M}$ and ${a}$ are the two defining parameters for the model. Again, ${h(r)=f(r)}$ automatically implies the regularity of the curvature scalars at the points ${\Phi_P^{(0)}\neq \frac{r_P}{2M}, d\neq 0}$. At the points where $\Phi_P^{(0)}=\frac{r_P}{2M}, d\neq 0$, the coefficients are given by equation \eqref{dymcoeffh}, and thus lead to a regular horizon.\par
The major difference between the Visser black hole and the models examined previously is that the metric functions \eqref{metrviss} are not analytic near the origin, i.e., the Taylor expansions of $h(r),f(r)$ do not equal the functions themselves. Nevertheless, ${h(r),f(r)}$ are infinitely differentiable, and hence one can Taylor approximate the functions near the origin
\begin{equation}\label{visszero}
    h(r)=f(r)=1+\mathcal{O}(r^3),
\end{equation}
where ${\mathcal{O}(r^3)}$ will include exponential corrections that tend to zero faster than any positive power of ${r}$. From \eqref{visszero}, one can immediately read off the coefficients of the deformation functions
\begin{equation}\label{exporvis}
\begin{split}
&\Phi_0^{(0)}=\Phi_0^{(1)}=\Phi_0^{(2)}=\Phi_0^{(3)}=0, \\ 
&\Psi_0^{(0)}=\Psi_0^{(1)}=\Psi_0^{(2)}=\Psi_0^{(3)}=0.
\end{split}
\end{equation}
As expected, the Visser model is regular at the origin. Further examination using the conditions from Sec.~\ref{sec:4.3} also confirms that the Visser model possesses a Minkowski core.\par
Finally, we use the additional regularity conditions derived in Sec.~\ref{sec:4.2} to investigate the regularity of geodesics. Similarly to the case of the Dymnikova black hole, ${\Phi(d)=\Psi(d)}$ and the necessary conditions for geodesic completeness are satisfied. The Visser model serves as a good example that EMDs are an effective way of describing black hole models which are not necessarily analytic everywhere.\par
\section{Conclusions}\label{Conclusion}

In this work, we investigated the regularity and geodesic completeness of static, spherically symmetric spacetimes within the Effective Metric Description. To do so, we generalized the EMD expansion around an arbitrary point. In this way both regions inside and outside the horizon can be parameterized in terms of the EMD deformation functions. This, in turn, allowed us to derive general conditions under which scalar curvature invariants remain finite across the entire spacetime, including at the origin. A novel addition beyond requiring curvature regularity is the study of geodesic completeness of the deformed metrics. By analyzing the effective potential governing the geodesics, we identified necessary constraints for the extendibility of null and timelike geodesics across the entire spacetime. 
The generality and utility of our results were illustrated by applying the EMD formalism to three representative cases: the singular Schwarzschild metric and the regular Dymnikova and Visser black holes. Our analysis is consistent with the regularity of the latter two and correctly reproduces the singularity of the Schwarzschild solution. Notably, we showed that even non-analytic models, such as the Visser black hole, can be successfully treated within our framework. Our results align with previous findings regarding the geometric and causal structure of regular black holes while offering new insights. For instance, we found that the regularity requires an even number of horizons, which agrees with earlier geometric arguments. Future work could explore the extension of our approach beyond static and spherically symmetric spacetimes, possibly incorporating rotating or collapsing scenarios. Potential applications of our results include the endpoint evolution of primordial black holes and the swampland program \cite{Palti:2019pca,Carr:2005zd}.  

\section*{Acknowledgements}
We thank Manuel Del Piano and Stefan Hohenegger for helpful discussions. The work of M.D.P., M.M., F.S., and V.V. is partially supported by the Carlsberg Foundation, grant CF22-0922.

\newpage
\appendix

\section{Invariant Definition of the Deformation Functions}\label{Appx:C}
In this appendix, we show explicitly how the deformation functions $\Psi(d)$ and $\Phi(d)$ can be defined in a generic set of coordinates $X$.\par
First, we find the Killing vectors ${\xi_t,\xi_i,i=1,2,3}$, which correspond to time translations and rotational invariance, respectively. The three vectors ${\xi_1,\xi_2,\xi_3}$ together form a basis of the $\mathfrak{so}(3)$ Lie algebra, and we can construct the quadratic invariant $L^2=\sum_i\xi_i\xi_i$. Subsequently, the “radial” direction can be defined as the direction perpendicular to all the Killing vectors ${\xi_t,\xi_i,i=1,2,3}$. This works because only two of the three Killing vectors $\xi_i$ associated with rotational symmetry are linearly independent. Once this is done, we can calculate the proper distance $d$ along the radial path away from the origin.\par
The deformation functions $\Psi(d),\Phi(d)$ can then be defined via coordinate invariants. The function ${\Psi(d)}$ is related to the norm of the timelike Killing vector:
\begin{equation}
    \Psi(d)=\frac{L}{2\sqrt{2}M}\left(1+\xi_t\cdot\xi_t\right).
\end{equation}
Once ${\Psi(d)}$ is known, the remaining function ${\Phi(d)}$ can be obtained by evaluating
\begin{equation}\label{retrphi}
    \Phi(d)=\frac{L}{2\sqrt{2}M}\left(1+2\kappa^2(\xi_t\cdot \xi_t)\left(\frac{dL}{d(\xi_t\cdot \xi_t)}\right)^2\right), \ \ \ \text{where} \ \ \kappa^2=\frac{1}{2}\nabla^\mu\xi_t^v\nabla_\mu\xi_{t,v}.
\end{equation}

In order to demonstrate how this works in practice, we carry out the calculations explicitly in Schwarzschild coordinates. The four Killing vectors can be chosen such that:
\begin{equation}
\begin{split}
&\xi_t=(1,0,0,0), \ \xi_1=(0,0,-\sin(\phi),-\cos(\phi)\cot(\theta)),\\
&\xi_2=(0,0,\cos(\phi),-\sin(\phi)\cot(\theta)), \ \xi_3=(0,0,0,1),\ 
\end{split}
\end{equation}
which gives $L^2=\sum_i\xi_i\xi_i=2r^2$. The radial direction is defined by the four-vector:
\begin{equation}
    (0,1,0,0).
\end{equation}
The function ${\Psi(d)}$ is then retrieved by calculating
\begin{equation}
    \frac{L}{2\sqrt{2}M}\left(1+\xi_t\cdot\xi_t\right)=\frac{r}{2M}\left(1-1+\frac{2M}{r}\Psi(d)\right)=\Psi(d).
\end{equation}
Similarly, the right-hand side of equation \eqref{retrphi} returns $\Phi(d)$:
\begin{equation}
    \kappa^2=\frac{f(r)}{4h(r)}\left(\frac{dh(r)}{dr}\right)^2\implies\frac{L}{2\sqrt{2}M}\left(1+2\kappa^2(\xi_t\cdot \xi_t)\left(\frac{dL}{d(\xi_t\cdot \xi_t)}\right)^2\right)=\frac{r}{2M}\left(1-f(r)\right)=\Phi(d).
\end{equation}
Since the above definitions are expressed in terms of coordinate invariants, they remain valid in any reference frame.

\newpage

\section{Expansion Coefficients}\label{Appx:A}
In this appendix, we provide explicit expressions for the expansion coefficients of the proper distance defined in Sec.~\ref{sec:3}:\par
\noindent Sec.~\ref{sec:3.1} (points ${\Phi_P^{(0)}\neq \frac{r_P}{2M}, d\neq 0}$):\par
\begin{equation}
    d_1=\frac{1}{\sqrt{|1-\frac{2M}{r_P}\Phi_P^{(0)}|}}, \ \ 
 \ d_2=\frac{M\left(r_P\Phi_P^{(1)}-\Phi_P^{(0)}\sqrt{|f(r_P)|}\right)}{2r_P^2|f(r_P)|f(r_P)}.
\end{equation}

\noindent Sec.~\ref{sec:3.2} (origin ${d=0}$):\par
{\bf Case 1: $\Phi_P^{(0)}=0$}\par
\begin{equation}\label{t3c1coeff}
    d_1: d_1=\frac{1}{\sqrt{\left|1-2Md_1\Phi_0^{(1)}\right|}}, \ \ \ d_2=\frac{1}{2}\frac{Md_1^2\Phi_0^{(2)}}{2d_1-4Md_1^2\Phi_0^{(1)}-M\Phi_0^{(1)}}.
\end{equation}\par
{\bf Case 2: $\Phi_P^{(0)}\neq 0$}\par
\begin{equation}
    d_1=\frac{2}{3\sqrt{|2M\Phi_0^{(0)}|}}, \ \ \ d_2=\frac{1}{10M\Phi_0^{(0)}\sqrt{|2M\Phi_0^{(0)}|}}, \ \ \ d_3=-\frac{\Phi_0^{(1)}}{18M\Phi_0^{(0)}|\Phi_0^{(0)}|}.
\end{equation}\par
\noindent Sec.~\ref{sec:3.3} (points $\Phi_P^{(0)}=\frac{r_P}{2M}, d\neq 0$):\par
{\bf Case 1: $\Phi_P^{(1)}=0$}\par
\begin{equation}
\begin{split}
    &d_1^{\pm} = \begin{cases}
            \pm\sqrt{\frac{1}{2M\Phi_P^{(2)}}\left(\pm1-\sqrt{1-16Mr_P\Phi_P^{(2)}}\right)}, \text{if $f(r)$ positive}\\
            \pm\sqrt{\frac{1}{2M\Phi_P^{(2)}}\left(\pm1+\sqrt{1+16Mr_P\Phi_P^{(2)}}\right)}, \text{if $f(r)$ negative}
           \end{cases}\\
    &d_2^{\pm}=\pm\frac{M(d_1^{\pm})^4\Phi_P^{(3)}}{12\mp 18M(d_1^{\pm})^2\Phi_P^2}
\end{split}
\end{equation}

{\bf Case 2: $\Phi_P^{(1)}\neq 0$}\par
\begin{equation}
    d_1^{\pm}=\pm\frac{3^{2/3}r_P^{1/3}}{2M^{1/3}\left|\Phi_P^{(1)}\right|^{1/3}}, \ \ \ d_2^{\pm}=\pm\frac{1}{8M\Phi_P^{(1)}}.
\end{equation}

\newpage
\section{Integrating the Proper Distance Equation}\label{Appx:B}
In this appendix, we comment on the technical details of the expansions for the proper distance carried out in Sec.~\ref{sec:3}. Although the computations from Sec.~\ref{sec:3.1} for ${\Phi_P^{(0)}\neq \frac{r_P}{2M}, d\neq 0}$ are straightforward, the ones from Sec.~\ref{sec:3.2} and Sec.~\ref{sec:3.3} involve two subtleties which deserve further discussion.\par
The first difficulty is encountered at the first order in equation \eqref{expeq0} when determining the leading coefficient $d_1$ of the proper distance expansion. If $\Phi_0^{(0)}=0$, the first coefficient
\begin{equation}
d_1=\frac{1}{\sqrt{\left|1-2Md_1\Phi_0^{(1)}\right|}}
\end{equation}
cannot be solved for exactly and must instead be given implicitly. Furthermore, the above equation for $d_1$ generally has more than one solution for a given value of $\Phi_0^{(1)}$. Fortunately, neither the regularity conditions nor the core classification depends on the value of $d_1$, and this ambiguity does not affect the subsequent results.\par
The second difficulty appears at higher orders in the expansion of equations \eqref{expeq0} (in the case $\Phi_0^{(0)}=0$) and \eqref{diffeq3} (in the cases $\Phi_P^{(0)}\neq 0$ or $\Phi_P^{(1)}\neq 0$), where a higher-order power of $\delta r$ in the proper distance expansion is not fully fixed by the lower order corrections. To illustrate how this problem arises, consider the expansion of \eqref{expeq0} to the higher order in $\delta r$ in the case $\Phi_P^{(0)}=0$. Using the first-order result, we can parameterize $d(r)\approx d(r_P)+d_1\delta r+d_2\delta r^j+...$, where $j$ and $d_2$ must be determined by matching the powers of $\delta r$ on both sides of the proper distance equation \eqref{propdistdiff}. This gives:
\begin{equation}\label{2ndorder}
\begin{split}
    &d_1+j d_2\delta r^{j-1}+...=\frac{1}{\sqrt{|1-2Md_1\Phi_0^{(1)}-2Md_2\Phi_0^{(1)}\delta r^{j-1}-Md_1^2\Phi_0^{(2)}\delta r+...|}}\\
    &\approx \frac{1}{\sqrt{\left|1-2Md_1\Phi_0^{(1)}\right|}}\left(1+\frac{M\Phi_0^{(1)}d_2}{1-2M\Phi_0^{(1)}d_1}\delta r^{j-1}+\frac{1}{2}\frac{M\Phi_0^{(2)}d_1^2}{1-2M\Phi_0^{(1)}d_1}\delta r+...\right)\\
    &\implies j d_2\delta r^{j-1}=\frac{M\Phi_0^{(1)}d_1d_2}{1-2M\Phi_0^{(1)}d_1}\delta r^{j-1} \ \ \text{if $1<j<2$},
\end{split}
\end{equation}
where we used the leading-order result for the coefficient $d_1$ in equation \eqref{t3c1coeff} to eliminate the constant terms.\par
The powers $\delta r^{j-1}$ cancel out as long as $1<j<2$, requiring only the preceding coefficients to match:
\begin{equation}\label{expl2order}
jd_2=\frac{M\Phi_0^{(1)}d_1d_2}{1-2M\Phi_0^{(1)}d_1}.
\end{equation}
The coefficient $d_2$ drops out of the equation as well, and hence, \eqref{expl2order} together with the first-order result \eqref{t3c1coeff} form a system of equations for $j$:
\begin{equation}\label{seteqj}
    j=\frac{M\Phi_0^{(1)}d_1}{1-2M\Phi_0^{(1)}d_1}, \ \ \ d_1=\frac{1}{\sqrt{\left|1-2Md_1\Phi_0^{(1)}\right|}}.
\end{equation}
It can be verified that the above system of equations does not have a solution with $j$ in the assumed range $1<j<2$. Therefore, the only possibility is for $j$ to be equal to $2$, such that the equations \eqref{seteqj} are modified by the next term ${\sim \delta r}$ in \eqref{2ndorder}. The solution to the resulting equation for $d_2$ is then given in \eqref{t3c1coeff}.\par 
Analogous issues in the expansions of $d(r)$ appear at higher orders in $\delta r$ at the points $d=0$ (the origin) and $\Phi_P^{(0)}=\frac{r_P}{2M}, d\neq 0$. The powers of $\delta r$ cancel out, and comparing the preceding coefficients leads to a system of equations for $j$. However, in both cases, the ambiguities are resolved when the lower-order coefficients $d_n$ are considered. 
\newpage

\sloppy

\newpage
\printbibliography[heading=bibintoc,title={References}]
\end{document}